\documentstyle[prl,aps,multicol,epsf]{revtex}

\begin{document}

\title{
 Ferromagnetism in multi--band Hubbard models: \\
 From weak to strong Coulomb repulsion
}

\author{
Karlo Penc$^1$\cite{st}\cite{empenc}, 
Hiroyuki Shiba$^2$\cite{emshiba}, 
Fr\'ed\'eric Mila$^3$\cite{emmila} 
and Takuya Tsukagoshi$^2$ \\
}
\address{\
$^1$Max-Planck-Institut f\"ur Physik komplexer Systeme, Bayreuther 
Str. 40, D-01187 Dresden, Germany \\
$^2$Department of Physics, Tokyo Institute of Technology, 
Oh-Okayama, Tokyo 152, Japan \\
$^3$Laboratoire de Physique Quantique, Univerisit\'e Paul 
Sabatier, 31062 Toulouse, France}

\date{\today} 
\maketitle

\begin{abstract}
We propose a new mechanism which can lead to ferromagnetism in Hubbard models
containing triangles with different on-site energies.
It is based on an effective Hamiltonian that we derive in the strong coupling
limit. Considering a one-dimensional realization of the model, we show
that in the quarter-filled, insulating case
the ground-state is actually ferromagnetic in a very large parameter
range going
from Tasaki's flat-band limit to the strong coupling limit of the effective
Hamiltonian. This result has been obtained using a variety of analytical and
numerical techniques. Finally, the same results are shown to apply away from
quarter-filling, in the metallic case.
\end{abstract}


\widetext
\begin{multicols}{2}
\narrowtext

\section{Introduction}

Notwithstanding its long history the study of itinerant ferromagnetism 
is known as one of the most difficult theoretical problems. 
At the early stage the problem of ferromagnetism was studied 
within the mean field approximation to the Coulomb interaction. 
However, in 1960's 
Kanamori,\cite{kanamori} Gutzwiller\cite{gutz} and Hubbard\cite{hubbard} 
independently studied the effect of 
electron correlations on metallic ferromagnetism, 
which is left out in the mean field approximation. 
According to them  correlation effects are so important that 
itinerant ferromagnetism is possible only in restricted 
situations: a large density of states at the Fermi energy is not 
sufficient to realize ferromagnetism. 
Nagaoka\cite{nagaoka} and Thouless\cite{thouless} pointed out a 
different aspect of ferromagnetism. Studying the motion 
of a hole (and a few holes) in strongly correlated electrons, 
Nagaoka and other researchers noticed that the possibility of 
ferromagnetism is a delicate issue depending on the lattice structure 
and the density of electrons.\cite{mh,edwards} 

Quite recently several exact results have been found for ferromagnetism 
in Hubbard and related models;\cite{vollhardt,yanagisawa} However 
those results are restricted to a special density of electrons. 
In this connection we note in particular 
that Tasaki has 
taken up the problem of ferromagnetism in a series of interesting 
papers\cite{tasaki1,tasaki2,tasaki3,kusakabe}, and has studied several 
new models, in which the relevant band is flat or almost flat. 
For such models he succeeded in proving rigorously
that the ground state is ferromagnetic when the electron density takes a 
particular value, 
namely 1/2  (quarter-filling), in which case  
the system is insulating. 

The purpose of the present work is to reexamine the same problem 
from a slightly different viewpoint. 
Our approach can provide a new way to look at the problem 
of ferromagnetism and can thereby examine the stability of ferromagnetism, 
at least in some cases, without any restriction to the electron density. 

Restricting ourselves for a while to one-dimensional systems, 
we will consider a model which interpolates between systems with large 
density of states and  Nagaoka--like ferromagnetism, and
we focus our attention on the following Hamiltonian: 
\begin{eqnarray}
{\cal H} &=&
  s\sum_{j,\sigma}\big(c_{j\sigma}^{\dagger}c_{j+1\sigma}
   +{\rm H.c.}\big) 
 - t\sum_{j,\sigma} 
     \big(d_{j-{1 \over 2}\sigma}^{\dagger}d_{j+{1 \over 2}\sigma}
    + {\rm H.c.}\big)
\nonumber \\
&&
 +t'\sum_j\big[(d_{j-{1 \over 2}\sigma}^{\dagger}
 +d_{j+{1 \over 2}\sigma}^{\dagger})c_{j\sigma}+{\rm H.c.}\big] 
 + \varepsilon_0\sum_j n_{j+{1 \over 2}\sigma}
\nonumber \\
&&
 + U \sum_j n_{j\uparrow}n_{j\downarrow}
 + U' \sum_j n_{j+{1 \over 2}\uparrow}n_{j+{1 \over 2}\downarrow},
\label{eqn:(1)}
\end{eqnarray}
where $s$ is assumed to be positive, while  $\varepsilon_0$ and $t$ are more
or less arbitrary and restrictions will be made when needed.
Index $j$ takes integer values, and sites with integer index we call `c' 
sites, while those with half--odd integer index `d' sites.
We consider cases where the number of
electrons $N_e$ is equal or less than the number of `c' sites 
($n \le \case{1}{2}$, where $n=N_e/L$ and $L$ is the total number of sites).
The ferromagnetic state has total spin $S=S_{\max}=\case{1}{2} N_e$.

\begin{figure*}
  \epsfxsize=7 truecm
  \centerline{\epsffile{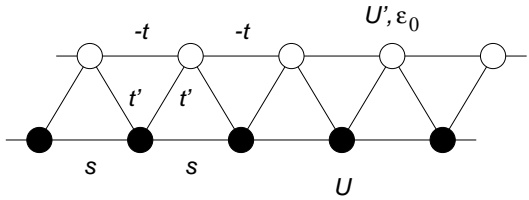}}
  \caption{
 Schematic representation on the 1D model. Solid circles stand for `c' sites,
and the open ones for `d' sites.
  }
  \label{fig:model}
\end{figure*}

This two band model, which is depicted in Fig.~\ref{fig:model}, has more
degrees of freedom than the regular Hubbard model and it should be able to 
exhibit more phases. Some special cases has already been studied in
literature. For $t=0$ and $\varepsilon_0=\varepsilon_{\rm FB}=t'^2/s-2 s$ the
dispersion of one of the bands becomes flat, and the ferromagnetism in the so
called flat band models has been discussed in the works of Mielke\cite{mielke}
and Tasaki.\cite{tasaki1}
Moreover, for $\varepsilon_0=0$, $U=U'$ and $t=-s$ the model is equivalent to
what M\"uller-Hartmann studied in the low density limit.\cite{mh2}  
Finally, for $\varepsilon_0=0$, $U=U'=+\infty$, $t=-s$ and $t' \ll t$ it was
discussed by Long and coworkers.\cite{Long}

 Let us also point out that by applying a particle--hole 
transformation to this model we get a Hamiltonian where the hopping 
amplitudes change sign, 
$\varepsilon_0$ becomes $-\varepsilon_0+U-U'$, the Coulomb interactions 
are unchanged, and the density transforms as $n \rightarrow 2-n$.

There are several reasons to believe that the ferromagnetic
ground state should be stable in Eq.~(\ref{eqn:(1)}) for a relatively wide 
range of parameters. 

\begin{figure*}
  \epsfxsize=8.5 truecm
  \centerline{\epsffile{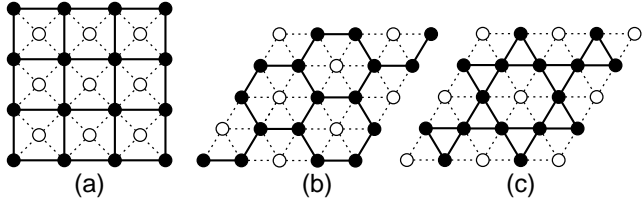}}
  \caption{
 Some possible realizations of the model in two dimensions: `c' sites 
form square (a), honeycomb (b) and kagome (c) lattice.
}
  \label{fig:model2d}
\end{figure*}

First, the basic building block is a triangle with two `c' sites and 
one `d' sites,
and the triangles are connected to form a chain (one dimensional model), or to
form some kind of two (see Fig.~\ref{fig:model2d}) or three dimensional 
lattice. Taking this triangle and putting 
two electrons on it, the triplet ground state is favored over the singlet for
a relatively large region of the hopping $t'$, $\varepsilon_0$ and interaction 
$U$. Especially, if $\varepsilon_0=t'^2/s -s$, then the high spin state is
realized for any finite value of $U$.

  Second,  as we already mentioned, for $\varepsilon_0=\varepsilon_{\rm FB}$
the lower band is flat, and it has been proved that the model is 
ferromagnetic for any $U>0$ at quarter filling ($n=1/2$). A nearly flat 
band also means large density of states near the Fermi level. 

  Third, anticipating the results of Sec.~\ref{sec:effHam}, we mention that
for $U=+\infty$ and large $\varepsilon_0$ the usual second
order process giving rise to effective antiferromagnetic coupling between
spins on neighboring sites in the case of the regular Hubbard model
is suppressed. Instead we have a third order
process with an effective exchange $\propto - s t'^2 /\varepsilon_0^2$, where
the sign can be adjusted by the hopping $s$ and it is ferromagnetic for
$s>0$. This is also valid away from quarter filling. The mechanism described 
is clearly of kinetic origin and resembles Nagaoka--effect.  
    
 These arguments raise the following questions on the model in 
Eq.~(\ref{eqn:(1)}):

  1) Does the high spin ground 
state survives if we put many triangles together? In other words, 
how robust is the Tasaki ferromagnetism against a
perturbation which destroys the flatness of the band? And to what extent 
is the strong coupling perturbational argument valid?

  2) Are there any cases in Eq.~(\ref{eqn:(1)}), where 
{\it metallic} ferromagnetism is realized? 

 \section{Summary of the results}  
 \label{sec:sum}    

\begin{figure*}
  \epsfxsize=8.5 truecm
  \centerline{\epsffile{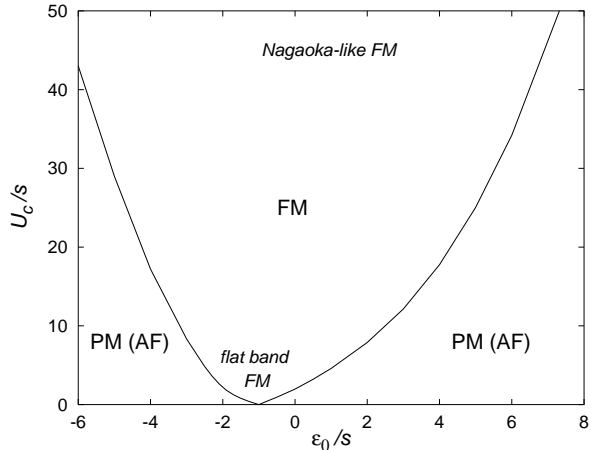}}
  \caption{
 Phase diagram for $t=0$, $s=t'=1$ and quarter filling. 
A flat band is realized for 
$\varepsilon_0=-1$ and for $U=+\infty$ the model is ferromagnetic for any
$\varepsilon_0$. For $U<U_c$ the system is paramagnetic, with strong
antiferromagnetic correlations. }
  \label{fig:sche}
\end{figure*}

To motivate 
the reader's curiosity, let us start by showing in Fig.~\ref{fig:sche}
the phase diagram we got in the insulating,
quarter-filled case. We see that ferromagnetism with 
a correct excitation spectrum turns out to be a robust
feature of this model and is realized in a fairly large parameter region. It
is not surprising if we recall that in some loose sense our model interpolates
between the Nagaoka like ferromagnetism of kinetic origin in the large-U limit 
and flat band ferromagnetism with large density of states at the Fermi level,
and that they both
favor the fully polarized, ferromagnetic ground state. This is very
encouraging since it shows that ferromagnetism can show up for cases that are 
more realistic
than the flat band model or Nagaoka effect. We emphasize
that near flat band models are multi--band systems, with the important
consequence that the interaction becomes momentum dependent ( and is not
constant as in the one--band Hubbard models) and this effective long range
interaction is essential to stabilize the ferromagnetic state for small
values of $U$. Furthermore the role of the triangles is not restricted to
generate effective ferromagnetic exchange, but also prevents
ferrimagnetism \cite{Lieb} (not being a bipartite lattice) and its importance
has been already recognized by Takahashi\cite{Takahashi} and 
Shastry {\it et al.}\cite{shastry}. Unfortunately, for the itinerant case the 
situation is not so clear. Although the phase diagram is similar to that of the
insulating case, the boundary effects due to low dimensionality of the model 
are enhanced and make the analysis difficult.
 
 Finally, let us remark that although we studied the one--dimensional 
model in details, we believe that similar conclusions would hold for 
higher dimensional models as well.

To reach the above conclusions we analyzed the model in Eq.~(\ref{eqn:(1)})
with different techniques in the appropriate limits:

\paragraph{Limit of $U=+\infty$ and $\varepsilon_0 \rightarrow \infty$, 
   any density.} 
  In this limit we use a canonical transformation to derive a low energy
  effective Hamiltonian and we show that ferromagnetic exchange is generated
  in the third order. 

\paragraph{Large $U$ and $\varepsilon_0$, quarter filling ($n=1/2$):}
 Rayleigh-Schr\"odinger perturbation theory allows us to derive a series
expansion for the dispersion of the spin flipped state up to 7th order. 
The series is then analyzed with the help of Pad\'e approximants and 
we determined the boundary of the local stability within the radius of 
convergence. In the lowest order the model is 
ferromagnetic for $U > U_c \approx s \varepsilon_0^2/t'^2$.

\paragraph{Small $U$ and nearly flat band, quarter filling.}
  For the cases when both $U$
and the width of the lower band is small comparable to the band gap, we
concentrate on the Hamiltonian restricted to states in the lower band.
 We find that there is a very nice 
Goldstone mode (magnon) with dispersion $\omega(q) = D q^2 + {\cal O}(q^4)$ 
for $q \rightarrow 0$,
where the spin stiffness $D$ can be calculated analytically. Furthermore,
the ferromagnetic state is stable for $U>U_c \propto |\rho|$, where $\rho$
measures the flatness of the band and is defined as 
$\rho = \varepsilon_0 - \varepsilon_{\rm FB}$.

\paragraph{Quarter filling, arbitrary $U$ and $\varepsilon_0$, 
  small size systems.} 
We diagonalize small clusters in the $S^z = S_{\rm max}-1$ (spin flipped)
subspace up to 20 sites and in the $S^z =0$ subspace up to 16 sites. The
data are consistent with the results obtained in the aforementioned limits,
and we fill the phase diagram for any $\varepsilon_0$. The global stability
seems to indicate that the transition is from ferromagnetic to singlet state,
except near the flat band where we do not have reliable data due to 
convergence problems. We find also that near $\varepsilon_0 =-2 s$ the system 
is very sensitive to boundary conditions due to orbital degeneracy of the 
ferromagnetic state.

\paragraph{$U=+\infty$ and large $\varepsilon_0$, less than quarter filling
($n<1/2$)}
 Using the factorized wave function, we determine the sign of the effective 
spin--spin interaction for different values of $s$, $t'$ and $t$. We discuss
the excitation spectrum of the spin flipped state for  
$J_{\rm eff} \ll s, t', t$, and show that in case of orbital degeneracy of 
the ferromagnetic state we need a finite value of $J_{\rm eff}$ to make the 
high spin state the lowest in energy.
 
\paragraph{Arbitrary $U$ and $\varepsilon_0$, 
  eighth filling ($n=1/4$), small size systems.} 
 From exact diagonalization of $L=12,16,20,24$ site systems in the 
$S^z = S_{\rm max}-1$ subspace with $N_e=L/4$ electrons we determine the
boundary of the local stability. Due to large finite size effect the 
results are not always conclusive. The effect of orbital degeneracy is 
enhanced compared to the insulating quarter filled system.

\section{Low-Energy Effective Hamiltonian 
         in the strong coupling limit }
 \label{sec:effHam}    

  In this section, we will derive an effective Hamiltonian 
when $t, t', s\ll \varepsilon_0, U, U'$.
In this limit, we have to distinguish between two cases  
 depending on the relative 
magnitude of $\varepsilon_0$ to $U$ and $U'$: $\varepsilon_0<U, U'$ 
and $U, U'<\varepsilon_0$. First we study the former case and we
 take the large--$U$ limit, {\it i.e.} 
$U, U' \to +\infty$. 
We mainly discuss the case where the electron density  
satisfies $n\le \case{1}{2}$. Since $\varepsilon_0$ is positive, 
only the lower band consisting mostly of the `c' component is filled. 

For $U, U'\to \infty$ the Hamiltonian is reduced to 
\begin{eqnarray}
{\cal H}&=&
 s\sum_{j,\sigma}
\big(\tilde c_{j\sigma}^{\dagger}\tilde c_{j+1\sigma}
+{\rm H.c.}\big) 
-t\sum_{j,\sigma}\big(\tilde d_{j-\case{1}{2}\sigma}^{\dagger}
\tilde d_{j+\case{1}{2}\sigma}+{\rm H.c.}\big)
 \nonumber \\
&& 
+t'\sum_{j,\sigma}\big[(\tilde d_{j-\case{1}{2}\sigma}^{\dagger}
+\tilde d_{j+\case{1}{2}\sigma}^{\dagger})\tilde c_{j\sigma}
+{\rm H.c.}\big]
+\varepsilon_0\sum_{j,\sigma} n_{j+\case{1}{2}\sigma} ,
 \nonumber \\
&& 
\label{eqn:(2)}
\end{eqnarray}
where $\tilde c_{j\sigma}$ and $\tilde d_{j-\case{1}{2}\sigma}$ are 
annihilation operators under the constraint of no double occupancy: 
$\tilde c_{j\sigma}=c_{j\sigma}(1-n_{j \bar\sigma})$, 
$\tilde d_{j-\case{1}{2}\sigma}=d_{j-\case{1}{2}\sigma}(1-n_{j-\case{1}{2}
\bar\sigma})$ and $\bar \sigma = -\sigma$. 
The creation operators are defined similarly. 

To derive an effective Hamiltonian to describe low-energy physics, 
we can treat the $t'\ll \varepsilon_0$ term as a perturbation 
inducing  transitions between the lower and upper bands. 
We use the canonical transformation to eliminate the off-diagonal 
processes due to 
\begin{equation}
{\cal H}'=t'\sum_{j\sigma}\big[(\tilde d_{j-\case{1}{2}\sigma}^{\dagger}
+\tilde d_{j+\case{1}{2}\sigma}^{\dagger})\tilde c_{j\sigma}+{\rm H.c.}\big]. 
\label{eqn:(3)}
\end{equation}
The other terms in Eq.~(\ref{eqn:(2)}) are denoted as ${\cal H}_0$. 
Since the canonical transformation is standard, we skip the details. 
Eliminating the first-order effect of $t'$, an effective Hamiltonian 
 up to the second order in $t'$ is obtained as follows: 
\end{multicols}
\widetext
\begin{eqnarray}
 \langle m'\vert {\cal H}_{\rm eff}\vert m\rangle
&=& \langle m\vert{\cal H}_0\vert m\rangle \delta_{mm'} 
+{1 \over 2}\sum_{n}\biggl({\langle m'\vert{\cal H}'\vert n\rangle 
\langle n\vert{\cal H}'\vert m\rangle \over E_m-E_n}
+{\langle m'\vert{\cal H}'\vert n\rangle 
\langle n\vert{\cal H}'\vert m\rangle \over E_{m'}-E_n}\biggr) ,
\label{eqn:(4)}
\end{eqnarray}
where $m, m', n$ are eigenstates of ${\cal H}_0$. 
The energy denominator $E_m-E_n$ can be written as 
$E_m-E_n=-\varepsilon_0+E_m'-E_n'$. 
Here $E_m'$ represents the part due to $t$ and $s$ 
in the eigenvalue $E_m$, {\it i.e.} ${\cal H}_0'\vert m\rangle
=E_m'\vert m\rangle$ where ${\cal H}_0'$ denotes ${\cal H}_0$ 
except for the $\varepsilon_0$ term. Since $\varepsilon_0$ is assumed 
to be much larger than $t$ and $s$, we can expand the energy 
denominators in the second term of Eq.~(\ref{eqn:(4)}) 
in terms of ${\cal H}_0'$ as
\begin{eqnarray}
\langle m'\vert {\cal H}_{\rm eff}\vert m\rangle
&=&\langle m\vert{\cal H}_0\vert m\rangle \delta_{mm'} 
-{1 \over \varepsilon_0}\langle m'\vert{\cal H}'{\cal H}'\vert m
\rangle 
+{1 \over 2\varepsilon_0^2}\Big(\langle m'\vert{\cal H}'
[{\cal H}_0', {\cal H}']\vert m\rangle+\langle m'\vert
[{\cal H}', {\cal H}_0']{\cal H}'\vert m\rangle\Big)  
\nonumber \\
&&
-{1 \over 2\varepsilon_0^3}\Big(\langle m'\vert{\cal H}'
[{\cal H}_0, [{\cal H}_0', {\cal H}']]\vert m\rangle+\langle m'
\vert[[{\cal H}', {\cal H}_0'], {\cal H}_0']{\cal H}'\vert m
\rangle\Big)  
+\cdots . \label{eqn:(6)}
\end{eqnarray}

\begin{multicols}{2}
\narrowtext

As mentioned before, we assume $n \leq \case{1}{2}$. 
Then only the lower band is filled. 
Having such states in mind, we evaluate each term in Eq.~(\ref{eqn:(6)}). 
The result is as follows:
\begin{eqnarray}
{\cal H}_{\rm eff}^{(1)}&=&s\sum_{j\sigma}\Big(\tilde 
c_{j\sigma}^{\dagger}
\tilde c_{j+1\sigma}+{\rm H.c.}) ,
\label{eqn:(7)}  
\\
{\cal H}_{\rm eff}^{(2)}&=&-{t'^2 \over \varepsilon_0}
\sum_{j\sigma}
\Big(2 n_{j\sigma}^{\dagger} 
+\tilde c_{j-1\sigma}^{\dagger}\tilde c_{j\sigma}
+\tilde c_{j+1\sigma}^{\dagger}\tilde c_{j\sigma}\Big), 
\label{eqn:(8)}  
\end{eqnarray}
and ${\cal H}_{\rm eff}^{(3)} = {\cal H}_{\rm eff}^{(3a)}
+{\cal H}_{\rm eff}^{(3b)}$ with  
\begin{eqnarray}
{\cal H}_{\rm eff}^{(3a)}&=&-{tt'^2 \over \varepsilon_0^2} 
\sum_{j\sigma}\Big[2 n_{j\sigma}
+2(\tilde c_{j-1\sigma}^{\dagger}
+\tilde c_{j+1\sigma}^{\dagger})\tilde c_{j\sigma}
\nonumber \\
&&+(\tilde c_{j-2\sigma}^{\dagger}
+\tilde c_{j+2\sigma}^{\dagger})\tilde c_{j\sigma}\Big], 
\label{eqn:(9)} 
\\
{\cal H}_{\rm eff}^{(3b)}
&=&-{s t'^2 \over \varepsilon_0^2}
\sum_{j\sigma}\Big[2\tilde c_{j\sigma}^{\dagger}
(\tilde c_{j-1\sigma}+\tilde c_{j+1\sigma}) 
\nonumber \\
&& +(n_{j-1\sigma}+n_{j+1\sigma})(1-n_{j\bar\sigma}) 
\nonumber \\
&& +(\tilde c_{j-1\sigma}^{\dagger}\tilde c_{j-1\bar\sigma}
+\tilde c_{j+1\sigma}^{\dagger}\tilde c_{j+1\bar\sigma})
\tilde c_{j\bar\sigma}^{\dagger}\tilde c_{j\sigma} 
\nonumber \\
&&  +\tilde c_{j-1\sigma}^{\dagger}(1-n_{j\bar\sigma})
\tilde c_{j+1\sigma}+\tilde c_{j+1\sigma}^{\dagger}
(1-n_{j\bar\sigma})\tilde c_{j-1\sigma} 
\nonumber \\
&& +(\tilde c_{j-1\sigma}^{\dagger}\tilde c_{j+1\bar\sigma}
+\tilde c_{j+1\sigma}^{\dagger}\tilde c_{j-1\bar\sigma})
\tilde c_{j\bar\sigma}^{\dagger}\tilde c_{j\sigma}\Big].  
\label{eqn:(10)}
\end{eqnarray}

\begin{figure*}
\epsfxsize=6.0 truecm
\centerline{\epsffile{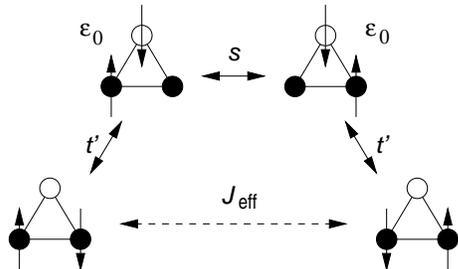}}
\caption{
 The third order process leading to ferromagnetic effective exchange $J_{\rm
eff}= 4 st'^2/\varepsilon_0^2$.
}
\label{fig:exch}
\end{figure*}

Although it looks complicated, Eqs.~(\ref{eqn:(7)})--(\ref{eqn:(10)}) 
become simple for $n=\case{1}{2}$, 
in which the states $\vert m\rangle$ are written as 
\begin{equation}
\vert m\rangle=\prod_{j} \tilde c_{j\sigma(j)}^{\dagger}
\vert {\rm vacuum}\rangle ,
\label{eqn:(11)}
\end{equation}
where the product is taken over all integer sites and where the quantum 
numbers $m$ just correspond to the set of spin indices ${\sigma(j)}$.
In fact it is easy to see that except for constant terms 
${\cal H}_{\rm eff}$ is reduced to 
\begin{equation}
 {\cal H}_{\rm eff}=-4s\Big({t' \over \varepsilon_0}\Big)^2\sum_{j}
 \Big({\bf S}_j\cdot{\bf S}_{j+1}-\case{1}{4}\Big),,
 \label{eqn:(12)}
\end{equation}
where ${\bf S}_j$ means a spin $\case{1}{2}$ at site $j$. 
Eq.~(\ref{eqn:(12)}) 
shows that the effective interaction is third order in the hopping; 
it is essentially 
a ferromagnetic Heisenberg model for positive $s$. 
Therefore the ground state is ferromagnetic with the 
maximum total spin.
The reason for this result is simple. 
The contribution in Eq.~(\ref{eqn:(12)}) comes from the ring exchange 
among 3 sites 
(a triangle formed with $j, j+1, j+\case{1}{2}$ sites, see 
Fig.~\ref{fig:exch}) with the hopping 
having the same sign, which is known to lead to a ferromagnetic 
exchange. 
Needless to say the system is insulating with a charge gap 
of $\sim\varepsilon_0$. Here we note that although 
Eq.~(\ref{eqn:(12)}) has been derived for the one-dimensional case, 
it is applicable to more general cases like the ones shown in 
Fig.~\ref{fig:model2d}. 
In that case the summation over $j$ should be interpreted 
as the one for all the nearest neighbor `c' sites.

So far we have assumed $U, U' \to \infty$. Relaxing this condition, 
let us take into account the lowest term in $1/U$. 
This is essentially the same as the one-band Hubbard model 
in the large-$U$ limit.\cite{ogata} The effective Hamiltonian 
which should be added to Eqs.~(\ref{eqn:(7)}-\ref{eqn:(10)}) is
\begin{eqnarray}
{\cal H}_{\rm eff}'&=& \frac{4 s^2}{U} \sum_{i}
\left({\bf S}_i\cdot {\bf S}_{i+1} - \case{1}{4}n_in_{i+1}\right) 
\nonumber\\
 &&+ \frac{s^2}{U} \sum_{i,\sigma}
      \Bigl(
         \tilde c^\dagger_{i,\sigma} 
         \tilde c^\dagger_{i+1,\bar\sigma} 
         \tilde c^{\phantom{\dagger}}_{i+1,\sigma} 
         \tilde c^{\phantom{\dagger}}_{i+2,\bar\sigma}
\nonumber\\
 &&      -  \tilde c^\dagger_{i,\sigma} 
         n^{\phantom{\dagger}}_{i+1,\bar\sigma} 
         \tilde c^{\phantom{\dagger}}_{i+2,\sigma} 
     + {\rm H.c.} \Bigr),  \label{eqn:(12a)} 
\end{eqnarray}
where $n_j=\sum_{\sigma}n_{j\sigma}$ is the density of electrons 
at site $j$, and  for quarter filling it simplifies to
\begin{equation}
{\cal H}_{\rm eff}' = \frac{4 s^2}{U} \sum_{i}
\left({\bf S}_i\cdot {\bf S}_{i+1} - \case{1}{4}\right) . 
  \label{eq:Ueff} 
\end{equation}
This Hamiltonian favors antiferromagnetic fluctuations and works against
ferromagnetism, so the final state will be decided by the competition 
between this Hamiltanion and that of Eq.~(\ref{eqn:(10)}).

 \section{ Stability of ferromagnetism at quarter-filling 
  ($\lowercase{n}=1/2$)}      
 \label{sec:quarter}    
   We now examine how robust ferromagnetism is 
at quarter-filling. We will look mainly at the stability against 
 one spin flip; this is the local stability 
condition. At some points we will mention the global stability as well.
 Henceforth we assume, for simplicity, $U'=U$ 
and $t=0$ in Eq.~(\ref{eqn:(1)}), in which case the topology of the model is 
the $\Delta$ chain. 

\subsection{ Large $U$ and $\varepsilon_0$}

  As we noted in the previous section, for $\varepsilon_0$ positive, and for 
$U$ and $\varepsilon_0$
much larger than the hopping amplitudes, the quarter filled system exhibits a
charge gap\cite{chargegap}
 $\Delta_c \approx \min(U,\varepsilon_0)$. 
Every `c' site is occupied by one electron, the excitations in the
charge sector are pushed up to high energies,
 and the only low lying excitations are coming from the 
spin degrees of freedom, which can be described in lowest order by 
the competition of Hamiltonians (\ref{eqn:(12)}) and (\ref{eq:Ueff}). 

To get reliable results in this limit, it is desirable to extend the
calculation of the effective spin Hamiltonian to higher order corrections.
Although in principle it can be done, see for example in 
Ref.~\onlinecite{MacDonald}, it is a very difficult task,
especially for a relatively complicated model as in Eq.~(\ref{eqn:(1)}). 
Therefore here we follow a slightly different approach, and we compute 
 the dispersion relation of the spin flipped state given by
\begin{equation}
   \varepsilon(k) = -K_1 \left[1-\cos(k)\right]-K_2 
\left[1-\cos(2k)\right]
 - \dots , 
  \label{eq:disp}
\end{equation}
 where the coefficients are related to the exchange integrals in the spin 
Hamiltonians. Were the spin Hamiltonian of the form 
$    J_1 \sum_{i} {\bf S}_i\cdot {\bf S}_{i+1} 
  + J_2 \sum_{i} {\bf S}_i\cdot {\bf S}_{i+2} + \dots $
then $K_1=J_1$, $K_2=J_2$ would hold. However, the canonical transformation
generates terms\cite{MacDonald} like 
$({\bf S}_i {\bf S}_{i'}) ({\bf S}_{j} {\bf S}_{j'})$ 
etc., and this makes the correspondence difficult.

  The series expansion for the coefficients in Eq.~(\ref{eq:disp}) 
can be obtained using the standard Rayleigh-Schr\"odinger perturbation theory.
As an initial state one picks up a state where all the `c' sites are occupied
with spin up electrons except for one site, where the spin points downwards,
furthermore all the `d' sites are empty. 
 Since the system is translationally invariant, momentum is a
good quantum number, and the problem is simplified as the  Hilbert space is 
spanned by   
states with given momentum $k$. In this way we avoid the $L/2$ fold
degeneracy of the ground state, as in every $k$ subspace the lowest
 energy state is unique. 
Now, through a number of virtual processes (hopping
to `d' sites, or hopping to already occupied sites), the spin down electron
will finally return to a `c' site. The distance $j$ which the down spin
electron has traveled will pick up a
phase factor $e^{i k j}$ and will contribute to
$K_j$. Although it does not look difficult, the technical realization of 
this calculation 
for higher orders is not straightforward. We used a variation of the method 
given by Barber and Duxbury,\cite{Barber} which is 
relatively simple. Recently, a more efficient method based on linked cluster
expansion become available due to Gelfand.\cite{Gelfand}

We have calculated the  coefficients $K_1, K_2, \dots$ up to 7th order, and
they are given in the Appendix. The lowest order contributions are
\begin{eqnarray}
  K_1^{(2)} &=& \frac{4s^2}{U}, \nonumber\\
  K_1^{(3)} &=& -\frac{4 s t'^2}{\varepsilon_0^2}
              -\frac{8 s t'^2}{U \varepsilon_0},   
\end{eqnarray}
in agreement with Eqs.~(\ref{eqn:(12)}) and (\ref{eq:Ueff}).

The first contribution to $K_2$ comes in 4th order,
\begin{eqnarray}
K_2^{(4)}&=&{ 4 s^4 \over U^3}, 
  \nonumber\\
K_2^{(5)}&=&
   -{16 s^3 t'^2 \over \varepsilon_0 U^3} 
   -{8 s^3 t'^2 \over \varepsilon_0^2 U^2}, 
\end{eqnarray}
and  $K_3$ appears first in 6th order (see Appendix for details).

\begin{figure*}
  \epsfxsize=8 truecm
  \centerline{\epsffile{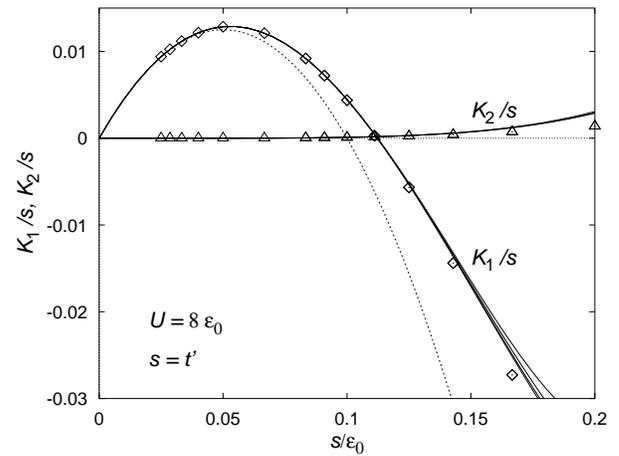}}
  \caption{
    The different Pad\'e approximants for $K_1$ and $K_2$ are shown for 
    $U/\varepsilon_0=8$. The series is convergent for 
    $s/\varepsilon_0 \sim 0.15$.
    The dashed line represents $K_1^{(2)}+K_1^{(3)}$.
    Note the excellent agreement with the values extracted 
    from exact diagonalization data of a 16
    site cluster (diamonds
    and triangles). The finite size effects are negligible.
  }
  \label{fig:j1j2}
\end{figure*}

\begin{figure*}
  \epsfxsize=8.6 truecm
  \centerline{\epsffile{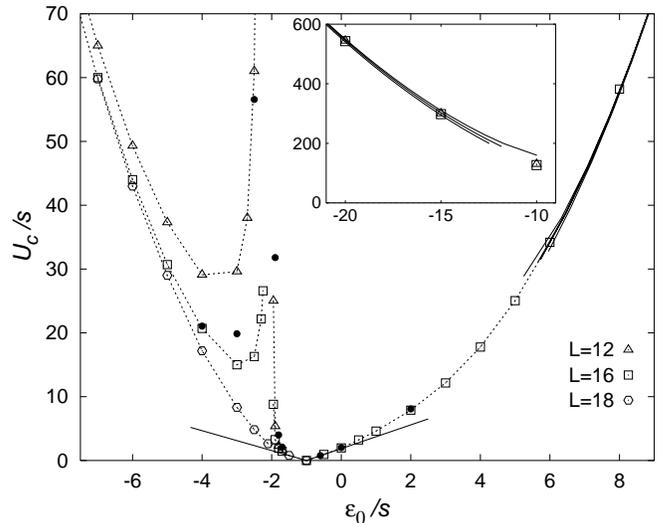}}
  \caption{
     Stability of the ferromagnetic state for $s=t'$ and $t=0$, quarter 
filling. Solid lines for large values
of $U_c$ come from Pad\'e approximants of the perturbational expansion (main
figure* and insert).
Solid straight lines are the approximation near the flat band. Opens symbols 
are the data for local stability for system sizes 12, 16 and 18 with periodic
boundary condition. The finite size effect is negligible apart from the
effect of degeneracy near $\varepsilon=-2s$ for 12 and 16 site systems. 
Solid points are global stability data of the 16 site cluster.
  }
  \label{fig:locsta}
\end{figure*}

To be more specific, let us now turn to the special case of $t'=s$.
The series can be analyzed with Pad\'e
approximants. A typical plot of $K_1$ as a
function of $s/\varepsilon_0$ is given in Fig.~\ref{fig:j1j2}. 
Fixing the ratio $U/\varepsilon_0$, we 
look for which value of $\varepsilon_0$ the coefficient $K_1$ changes sign. 
It is enough to consider 
$K_1$ since $|K_2|\ll|K_1|$ holds near the transition.
Changing the ratio of $U/\varepsilon_0$, we are able 
to determine the
region where ferromagnetism is locally stable. 
The result is shown in Fig.~\ref{fig:locsta}. 
For $s\neq t'$ there are some quantitative changes in the location 
of the boundary, which can be already recognized from $K_1^{(2)}$ and 
$K_1^{(3)}$ yielding $U_c \approx s \varepsilon_0^2/t'^2$.

 A similar calculation can be carried over 
for $U,-\varepsilon_0 \gg s,t'$. 
Then the electrons sit on `d' sites and for  $t=0$ the first
contributions to $\bar K_1$ come in fourth order:
\begin{eqnarray}
\bar K_1^{(4)}&=& \frac{4 \tilde t^4}{\bar\varepsilon_0^2 U } 
      + \frac{8 \tilde t^4}{\bar\varepsilon_0^2(2\bar\varepsilon_0 + U)}, 
  \nonumber\\
\bar K_1^{(5)}&=&  
     - \frac{8\tilde t^5}{\bar\varepsilon_0^4} 
     - \frac{16 \tilde t^5}{ \bar\varepsilon_0^3 U} 
     - \frac{32 \tilde t^5}{ \bar\varepsilon_0^3 (2\bar\varepsilon_0 + U)}. 
\end{eqnarray}
where we concentrated on $s=t'=\tilde t$ case, 
$\bar \varepsilon_0 = - \varepsilon_0$ and the coefficients are denoted 
$\bar K_j$.
For higher order correction up to 7th order we refer the 
reader to the Appendix.
The series is convergent for
$|\varepsilon_0|/\tilde t \gtrsim 15$.

\subsection{ In the vicinity of the flat band}      

 Our model is equivalent to Tasaki's flat band model\cite{tasaki1} 
for $\varepsilon_0 =
\varepsilon_{\rm FB}= t^{\prime 2}/s -2 s $, where he has
rigorously proven that the model is 
 ferromagnetic for arbitrarily small value of interaction. 
In this section we restrict $\varepsilon_0$ to be close 
to $\varepsilon_{\rm FB}$, and we introduce a quantity $\rho$ which 
measures the perturbation from the flat band as 
$\rho =\varepsilon_0-\varepsilon_{\rm FB}$. 
In a series of recent papers\cite{tasaki2} Tasaki extended the existence of 
ferromagnetism for such perturbed models, reaching the result that 
ferromagnetism is stable for the nearly flat bands as well for $U>U_c$,
where $U_c$ is roughly proportional to the perturbation strength, 
in our case $\rho$.  
His treatment is rather involved and we are looking for a
simpler description.
 
Let us consider the model Hamiltonian in $k$ space:
\begin{eqnarray}
  {\cal H} &=& 
   2 s \sum_{k,\sigma} 
      \cos k  c^\dagger_{k\sigma} c^{\phantom{\dagger}}_{k\sigma} 
    +\varepsilon_0 \sum_{k,\sigma}
      d^\dagger_{k\sigma} d^{\phantom{\dagger}}_{k\sigma}
  \nonumber\\
   &&
   + 2 t' \sum_{k,\sigma} 
      \cos \frac{k}{2} 
    \left( c^\dagger_{k\sigma} d^{\phantom{\dagger}}_{k\sigma}
         + d^\dagger_{k\sigma} c^{\phantom{\dagger}}_{k\sigma}
    \right)
  \nonumber\\
   &&
   \!\!+ \frac{2U}{L} \! \! 
    \sum_{k_1,k_2,k_3,k_4} \!\!  
    \left(  
       c^\dagger_{k_1\uparrow}
       c^\dagger_{k_2\downarrow}
       c^{\phantom{\dagger}}_{k_3\downarrow}
       c^{\phantom{\dagger}}_{k_4\uparrow}
   \! +\!  d^\dagger_{k_1\uparrow}
       d^\dagger_{k_2\downarrow}
       d^{\phantom{\dagger}}_{k_3\downarrow}
       d^{\phantom{\dagger}}_{k_4\uparrow}
    \right),
  \nonumber\\
\end{eqnarray}
where in the summation $k_1+k_2-k_3-k_4=G$ should hold, $G=0,\pm 2 \pi$ 
is a reciprocical lattice vector. In the noninteracting case 
($U=0$) it describes two bands where the lowest band is nearly flat (the
bandwidth is $\propto \rho$), while the upper band is dispersive. The two
bands are separated by a band gap 
$ \Delta_{\rm{band}}= 2s + \varepsilon_0 + {\cal O}(\rho)$. 
We expect, as already noted by Kusakabe and Aoki,\cite{kusakabe} 
that for small values of $U$ the essential physics is now going on
in the lower band. To construct an effective Hamiltonian, 
we first diagonalize the hopping part of the Hamiltonian 
by the following canonical transformation
 (here we choose $t'>0$ for convenience, as the model 
does not depend on the
sign of $t'$):
\begin{eqnarray}
  c^\dagger_{k\sigma} &=& 
   \phantom{-} \cos \alpha_k a^\dagger_{k\sigma} 
                 + \sin \alpha_k b^\dagger_{k\sigma} ,
\nonumber\\
    d^\dagger_{k\sigma} &=&-\sin \alpha_k a^\dagger_{k\sigma} 
                 + \cos \alpha_k b^\dagger_{k\sigma} ,
\end{eqnarray}
where $a^\dagger_{k,\sigma}$ and $b^\dagger_{k,\sigma}$ are 
the creation 
operators of electrons on the lower and upper band, respectively. 
$\alpha_k$ is determined via
\begin{equation}
  \tan 2 \alpha_k = \frac{4 t' \cos(k/2)}{\varepsilon_0 - 2 s \cos(k)},
\end{equation}
and is a continuous function 
of the momentum $k$, so that
it is between 0 and $\pi/2$ for momenta in the Brillouin zone; 
furthermore $\alpha_{\pm \pi} = 0$  for 
$\varepsilon_0>-2s$ and $\alpha_{\pm \pi} = \case{\pi}{2}$ for 
$\varepsilon_0<-2s$. 
It is convenient to extend its definition
outside the Brillouin zone so that 
$\alpha_{ 2 \pi + k} = -\alpha_{k} $
for any $k$ in case of $\varepsilon_0>-2s$ and
$\alpha_{ 2 \pi + k} = \pi -\alpha_{k} $ 
for $\varepsilon_0<-2s$ .
The band dispersion is then given by
\begin{equation}
 \varepsilon_{a,b} (k)  = 
  \frac{ \varepsilon_0}{2} + s \cos k 
    \mp \sqrt{
       \left(\frac{\varepsilon_0}{2} - s \cos k \right)^2
        +4 t'^2 \cos^2 \frac{k}{2}}; 
\end{equation}
the effective Hamiltonian in the lower band reads
\begin{eqnarray}
  H &=& \sum_{k,\sigma} \varepsilon_a(k) a^\dagger_{k\sigma}
a_{k\sigma}
      \nonumber\\
     && + \sum_{k_1,k_2,k_3,k_4} 
     V(k_1,k_2,k_3,k_4) 
     a^\dagger_{k_1\uparrow}a^\dagger_{k_2\downarrow}
     a_{k_3\downarrow}a_{k_4\uparrow} ,
  \label{eq:Heffrho}
\end{eqnarray}
where the interaction is defined as
\begin{eqnarray}
  V(k_1,k_2,k_3,k_4) &=& \frac{2U}{L} 
  \Bigl(
   \cos \alpha_{k_1} \cos \alpha_{k_2}\cos \alpha_{k_3} 
\cos \alpha_{k_4} 
\nonumber\\
&& \!\!\!\!\!\!+ e^{i G /2} 
    \sin \alpha_{k_1} \sin \alpha_{k_2}\sin \alpha_{k_3} 
\sin \alpha_{k_4}
  \Bigr)  
  \label{eq:vdef} 
\end{eqnarray}
and $k_1+k_2-k_3-k_4=G$ holds.
The factor $e^{i G/2}$ is $1$ 
for normal scattering ($G=0$) and
$-1$ for umklapp scattering ($G=2\pi$), and it can be conveniently taken care 
of using the definition of $\alpha_{k}$ for $k$ values 
outside the Brillouin zone.

Here we neglected terms which are coming from the scattering on the 
states in the upper
band\cite{chargegap} and they are of the order of $U^2 / \Delta_{\rm{band}}$, 
 so that the approximation is valid for 
$U\ll \Delta_{\rm{band}}$ .

We now concentrate on the excitation spectrum 
of the spin flipped states 
of momentum q which are defined as 
\begin{equation}
|\chi(q,\nu) \rangle = \sum_p f_p^\nu(q) 
a^\dagger_{p+q,\downarrow} a_{p,\uparrow}
 | {\rm FP} \rangle , 
\end{equation}
where $|{\rm FP} \rangle $ is the fully 
polarized state with
all the spins up (i.e. $S=S_{\rm max}=L/2$), $q$ is measured from 
the momentum of  $|{\rm FP} \rangle $ and $\nu$ is some quantum number. 
Since the lower band is filled, these states span the whole reduced Hilbert 
space and no approximation is made in solving the
effective Hamiltonian (\ref{eq:Heffrho}).

 The dispersion of the spin flipped state 
$\omega(q)$ is given by the eigenvalue problem\cite{kusakabe}
 \begin{equation}
   \sum_{p'} M_{p,p'}(q) f_{p'}^\nu(q) = \omega_\nu(q) f_p^\nu(q),
  \label{eq:mfof}
 \end{equation}
where the matrix is given by 
\begin{eqnarray}
 M_{p,p'}(q) &=& 
  \left[
     \varepsilon_a(p+q) + \Delta(p+q)-\varepsilon_a(p) 
  \right] \delta_{p,p'} \nonumber\\
  &&- V(p,p'+q,p+q,p').
  \label{eq:mev}
\end{eqnarray}
$\Delta(p) = \sum_{k} V(k,p,p,k)$ is the Stoner gap.
The size of the matrix is $(L/2)\times (L/2)$ and can be
easily diagonalized numerically for large system sizes.

\begin{figure*}
  \epsfxsize=8.5 truecm
  \centerline{\epsffile{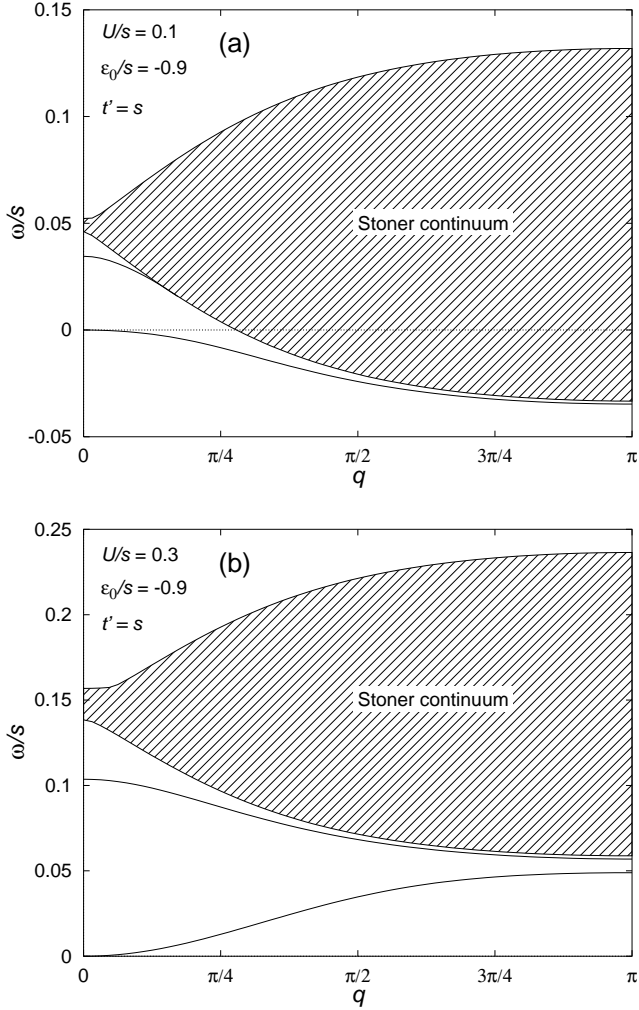}}
  \caption{
     The excitation spectra in the thermodynamic limit for some selected
cases. (a) For small values of $U$ the fully polarized state is not stable,
but already we can see a well defined magnon, which is very close to 
the particle-hole excitation continuum for $q=\pi$. (b) For larger values 
of $U$ we can see a normal ferromagnetic magnon dispersion relation, 
with the Stoner continuum
pushed up to higher energies.
  }
  \label{fig:stoner}
\end{figure*}

\begin{figure*}
  \epsfxsize=8.5 truecm
  \centerline{\epsffile{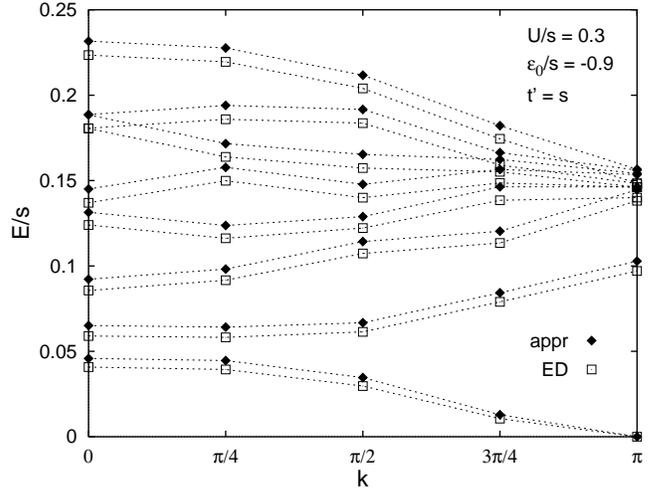}}
  \caption{
     The comparison of the excitation spectrum obtained from matrix 
$M_{p,p'}(q)$ (solid diamonds)
  to exact diagolization of a 16 site cluster (open squares). 
Although $U/s=0.3$ is already 
comparable to the band gap, the errors are very small. 
The discrepancy is due interband
 scattering and it scales as 
$U^2$. Here $k=\pi-q$, where $\pi$ is the momentum of $|{\rm FP}\rangle$.
  }
  \label{fig:edsto}
\end{figure*}

 A typical plot of the excitation spectra is given 
in Figs.~\ref{fig:stoner} and ~\ref{fig:edsto}. There we can see that a 
low energy magnon (Goldstone mode) 
and an optical magnon (with spin oscillation within the cell) 
emerge from the Stoner continuum. 
For small values of $q$ the magnon has a dispersion 
$\omega_0(q) = D q^2$.

Now, what are the conditions for local stability? First of all,
to have a  dispersion relation appropriate for ferromagnets, 
stiffness $D$ should be positive. The interaction
strength where $D$ changes sign we call $U_D$.
However, this is not enough, as it can happen that although $D>0$, the
Stoner continuum pushes down the spin wave mode at $q=\pi$. Therefore 
the decisive criterion for local stability is that all the excitation energies
are positive (i.e. all the energies are above the energy of the fully 
polarized state), and this will define $U_c$. 

 \subsubsection{Calculation of the $U_D$}
 Clearly, $D$ is given as 
$\frac{1}{2}\partial^2 \omega_0(q)/\partial q^2|_{q=0}$ and it can be
calculated from Eq.~(\ref{eq:mev}) by using Hellman-Feynman rule 
(it will give
$D_1$) and second
order perturbation theory ($D_2$), so that $D = D_1 + D_2$, where
\begin{eqnarray} 
 D_1 &=& \frac{1}{2}
   \Bigl.
     \langle \chi(0,0) |
       \frac{ \partial^2 M(q)}{ \partial q^2}
    | \chi(0,0) \rangle  \Bigr|_{q=0}, \nonumber\\
 D_2 &=&   
 - \sum_{\nu\neq 0}
    \frac{1}{\omega_\nu - \omega_0}
    \Bigl.
    \bigl|
      \langle \chi(0,\nu) |
      \frac{\partial M(q)}{\partial q} |  \chi(0,0) \rangle
    \bigr|^2
    \Bigr|_{q=0} .   
\end{eqnarray} 
Here $| \chi(0,0) \rangle$ is the ground state wave function 
for momentum $q=0$, which is nothing else but $S^-|{\rm FP}\rangle$ 
and is trivially given by
$f^0_{p}(0) = \sqrt{2/L}$. The  $|\chi(0,\nu)\rangle$ are excited
states. Since by definition $D_2$ is negative and 
 works against ferromagnetism, sufficiently large $D_1$ is needed to
compensate $D_2$ and to give $D>0$.

Nonzero contribution to $D_1$ will come from the off-diagonal 
matrix elements
of $M_{p,p'}$:
\begin{equation}
 D_1 = -\frac{1}{2}
  \int_{-\pi}^{\pi} \frac{dp}{2\pi} \int_{-\pi}^{\pi} \frac{dp'}{2\pi}  
      \Bigr.
       \frac{ \partial^2} { \partial q^2} V(p,p'+q,p+q,p')
      \Bigr|_{q=0}, 
\end{equation}
since the diagonal part vanishes being 
an integral of the derivative of periodic function.
Using the special form of interaction like in Eq.~(\ref{eq:vdef}) and 
carrying out partial integration, one can show that 
this quantity is positive, unless $V$ is a constant. In other words the
momentum dependence of the interaction is crucial to stabilize the 
ferromagnetic state.

 Let us now turn to $D_2$. In our case
the nonvanishing matrix elements come from odd-momentum 
excited states
labeled by momentum $\tilde p$, where
$f_{p}^{\tilde p}(0) = (\delta_{p,\tilde p} 
- \delta_{p,-\tilde p})/\sqrt{2}$
 with energies $\omega_\nu-\omega_0=\Delta(\tilde p)$. Thus we obtain 
\begin{eqnarray} 
  D_2&=& 
  - \frac{1}{4}\frac{2}{L}\sum_{\tilde p}
    \frac{1}{\Delta(\tilde p)}
    \left[
        \Bigl.  
        \sum_{p} \frac{\partial}{\partial q}
          \left(
             M_{p,\tilde p} - M_{-p,\tilde p}
          \right)
      \Bigr|_{q=0} 
    \right]^2
    \nonumber\\    
  &=& 
  - \frac{1}{4}\int_{-\pi}^{\pi}
     \frac{d \tilde p}{2 \pi}
     \frac{1}{\Delta( \tilde p)}
    \left[
      2 \frac{\partial \varepsilon( \tilde p)}{\partial \tilde p}
      + \frac{\partial \Delta(\tilde p)}{\partial \tilde p}
    \right]^2 .
  \label{eq:d2f}
 \end{eqnarray} 
The divergence of $D_2$ for small values of $U$ is the
signature that the dispersion becomes linear in $q$ for $U=0$ and the result
is non--perturbative in $U$.

To make the calculation simple we expand everything in powers of the 
small parameter $\rho$:
\begin{eqnarray}
   \cos \alpha_k &=& \frac{ t' }{ \sqrt{t'^2 
+ 4 s^2\cos^2(k/2)}} + {\cal O}(\rho), 
  \nonumber\\
   \sin \alpha_k &=& 
      \frac{ 2 s \cos(k/2) }{\sqrt{t'^2 + 4 s^2\cos^2(k/2)}}
+{\cal O}(\rho), 
   \nonumber\\
     \varepsilon(k) &=& -2 + (1-\cos^2 \alpha_k) \rho +{\cal O}(\rho^2). 
\end{eqnarray}
The following integrals are useful for our purpose:
\begin{eqnarray}
  I_{c} &=& \int_{-\pi}^{\pi} \frac{dk}{2 \pi} \cos^2 \alpha_k 
        = \frac{t'}{\sqrt{t'^2+4s^2}}, \nonumber\\
  I_{s} &=& \int_{-\pi}^{\pi} \frac{dk}{2 \pi} \sin^2 \alpha_k 
        = 1-I_c, \nonumber\\
  I'_{c} &=& \int_{-\pi}^{\pi} \frac{dk}{2 \pi} 
              \left(\frac{d\cos \alpha_k}{dk}\right)^2 
        = \frac{s^4}{2 t' (t'^2+4s^2)^{3/2}}, \nonumber\\
  I'_{s} &=& 
       \int_{-\pi}^{\pi} \frac{dk}{2 \pi} 
              \left(\frac{d\sin \alpha_k}{dk}\right)^2 
        = \frac{s^2}{2 t' \sqrt{t'^2+4s^2}} -I'_{c}.
\end{eqnarray}
With the help of these integrals, we can 
write the Stoner gap as
\begin{equation}
  \Delta(k) = 
   U \left( 
        I_c \cos^2 \alpha_k 
      + I_s \sin^2 \alpha_k 
     \right ).
\end{equation}
Finally, stiffness $D$ is the sum of the $D_1$ and $D_2$:
\begin{eqnarray}
  D_1 &=& U (I_c I'_c + I_s I'_s), \nonumber\\
  D_2 &=& - \frac{C}{U}
     \left[-2 \rho + (2 I_c-1) U\right]^2,  
\end{eqnarray}
where 
\begin{equation}
  C=  \int_{-\pi}^{\pi} \frac{dk}{2\pi} 
  \frac{ \cos^2 \alpha_k}{I_c \cos^2 \alpha_k + I_s \sin^2 \alpha_k}
  \left(\frac{\partial \cos \alpha_k}{\partial k}\right)^2.
\end{equation}
 Although the coefficient $C$ can be given 
in a closed form, 
it is rather complicated and we give its value 
for some selected values of $t'/s$ in Table.~\ref{tab:1}, together 
with the values of $U$ where $D$ changes sign ($U_D$). 
Note that the equation we solve is a quadratic one and 
it gives two solutions: one is for $\rho>0$ 
and the other one is for $\rho<0$.

In the limit $t'\gg s$ we get
\begin{equation}
  D= \left(U+2\rho - 2 \frac{\rho^2}{U} \right)\left( \frac{s}{t'}\right)^4
\end{equation}
for small $U$ and $\rho$,
which is in agreement with Tasaki's result.\cite{tasaki2}
\narrowtext
\subsubsection{Calculation of $U_c$}

 From numerical diagonalization of the matrix 
$M_{p,p'}$ we see that the other local  minimum 
of the acoustic magnon branch is at
$q=\pi$. Unfortunately we do not know how to get $U_c$, therefore
we have to solve the eigenvalue problem of Eq.~(\ref{eq:mfof}) numerically
and check for which values of $U$ $\omega(q=\pi)>0$ holds.  
 From this we obtain $U_c$. Again, we have to distinguish between 
$\rho>0$ and $\rho<0$. The results are presented in Table.~\ref{tab:1}.

We also learn that for small $t'$ a very good estimate of $U_c$ can
be obtained by using only the diagonal elements 
of matrix $M_{p,p'}(\pi)$ applying the Stoner criterion.
Namely, the spin 
down electron occupies states in an effective band 
$\varepsilon_\downarrow(p) = \varepsilon_a(p) + \Delta(p)$ 
due to the background of spin up electrons.
This band
is higher in energy than the $\varepsilon_\uparrow(p)=\varepsilon_a(p)$ 
of the spin up electrons. 
According to Stoner's theory, when lowering  $U$ the Stoner gap 
decreases, and at some point we start to fill the band of spin down electrons. 
If we neglect the off diagonal matrix elements describing the interaction 
between the hole in the spin up band and the spin down electron, this happens
exactly when the energies of the two bands start to overlap. For $\rho>0$
the minimum of $\varepsilon_\downarrow(p)$ is at $p=\pi$ and the maximum of 
$\varepsilon_\uparrow(p)$ is at $p=0$, so that the criterion is  
$\varepsilon_\downarrow(\pi)=\varepsilon_\uparrow(0)$. This  will give us 
$U_c^{\rm app} = 4 \rho / (t' \sqrt{t'^2 + 4s^2})$;
for $\rho<0$ the corresponding equation is
 $\varepsilon_\downarrow(0)=\varepsilon_\uparrow(\pi)$.

For $t'\gg s$, on the other hand, $U_D$ and $U_c$ are
 very close to each
other. This is due to the fact that the excited states 
in this case are far
from the acoustic magnon branch and their interference is small for values of
$U$ close to $U_c$. This was also predicted by Tasaki\cite{tasaki2} in the 
sense
that lower and upper bound on the dispersion relation are very close to each 
other in this case.

\begin{figure*}
  \epsfxsize=8.5 truecm
  \centerline{\epsffile{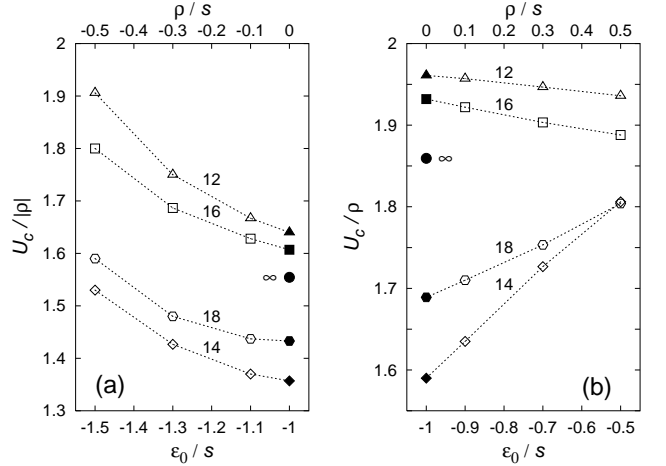}}
  \caption{    
  $U_c/|\rho|$ calculated from exact diagonalization (open symbols) and 
from  diagonalization of matrix $M_{p,p'}(\pi)$ (solid symbols) for system
sizes $L=12,14,16$ and 18 with periodic boundary conditions, $t'=s$ and $t=0$.
 The solid circle stands for the value of $U_c$ in
the thermodynamic limit $L\rightarrow \infty$. 
  }
  \label{fig:eom1}
\end{figure*}

We have compared these results  with the exact diagonalization
results for the stability of the spin flipped state in Fig.~\ref{fig:eom1}
for $t'=s$ and we found that $U_c$ increases linearly with $\rho$
 even for relatively large values of $\rho$ ($\approx 0.5 s$)for $\rho>0$,
whereas for
$\rho<0$ the linearity holds only for small values of $\rho$.

\subsection{The orbital degenaracy at $\varepsilon_0=-2s$}

\begin{figure*}
  \epsfxsize=8.5 truecm
  \centerline{\epsffile{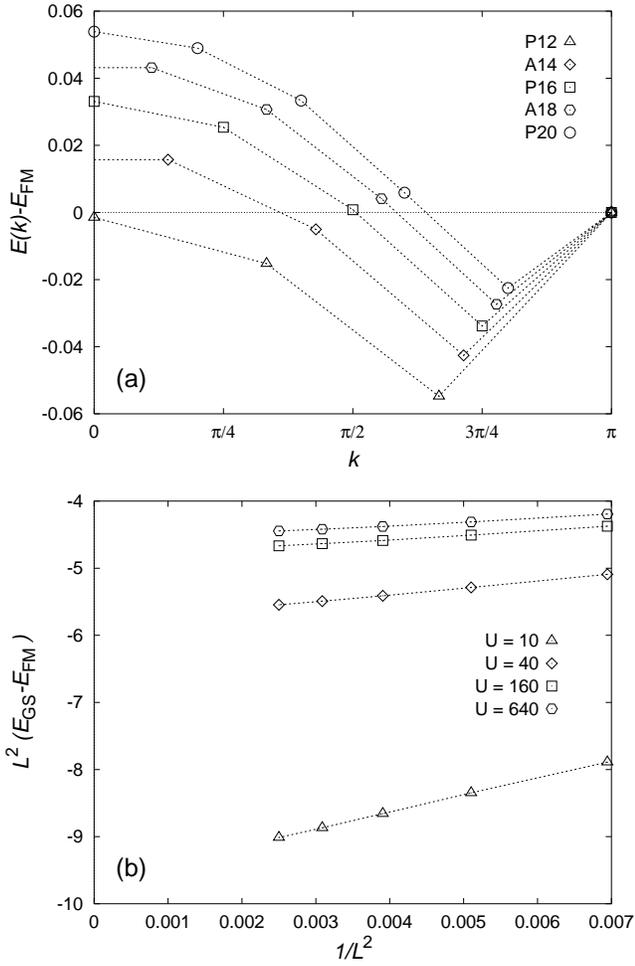}}
  \caption{ (a) The dispersion of the lowest energy spin flipped state 
for $t'=s=1$, 
$\varepsilon_0=-1$ and $t=0$ from exact diagonalization of 12,16, and
20 site clusters with periodic (P) and $L=14$ and 18 with antiperiodic 
(A) boundary
conditions. Although the $S_{\rm max}$ state at $k=\pi$ is higher in energy,
the dispersion has clearly ferromagnetic $k$ dependence. 
(b) The finite size scaling of the 
energy difference, $E_{\rm GS}$ is the lowest energy in the 
$S^z=S_{\rm max} -1$ subspace.   
  }
  \label{fig:deg}
\end{figure*}

  A very interesting point of our phase diagram is at 
$\varepsilon_0=-2 s$, where the upper and lower bands touch. Then, 
depending on the system size 
($L=8$, 12, 16, 20 ... with periodic boundary conditions or 
 $L=6$, 10, 14, 18 ... with antiperiodic boundary conditions), 
the fully polarized ground 
state can be degenerate.  This orbital degeneracy seems not to 
favor the highest spin states, as it was noticed e.g. in the case
of Nagaoka ferromagnetism by S\"ut\H o.\cite{Suto}
To show this explicitly, we calculated the energy dispersion in the 
$S^z =S_{\rm max}-1$ subspace (see Fig.~\ref{fig:deg}) and find that the
difference between the energy  $E$ of the groundstate in the $S_{\rm max}-1$ 
sector and the energy $E_{\rm FM}$ of the ferromagnetic state  satisfies the
following finite size scaling:
\begin{equation}
  E-E_{\rm FM} = \frac{C}{L^2} + {\cal O}(1/L^4),
\end{equation}
with $C<0$, thus for any 
finite system the ground state has a total spin less than $S_{\rm max}$.

It seems that this strange behavior is constrained to $\varepsilon_0 = -2s$ 
and away from it the $S_{\rm max}$ state is the ground state in the
thermodynamic limit for $U>U_c$. To clarify this issue more precisely 
further work is needed.

\section{Ferromagnetism in the metallic case ($\lowercase{n}<1/2$)}
 \label{sec:metallic}    

Here we are mainly concerned with the case of large $U$ and $\varepsilon_0$.
A nice feature of the effective Hamiltonian is that 
as far as the one-dimensional case is concerned, 
one can also discuss the metallic case corresponding to $n<1/2$. 
(A different way to show the stability of ferromagnetic 
state is given in Ref.~\onlinecite{tasakiiti}.)
Actually the metallic case is more interesting 
from a physical point of view than the insulating state. 
For $n<1/2$ all the terms in Eqs.~(\ref{eqn:(7)})--(\ref{eqn:(10)}) 
contribute. Although it looks complicated, the effective Hamiltonian 
in Eqs.~(\ref{eqn:(7)})--(\ref{eqn:(10)}) is similar to the $t-J$ model. 
Noting that $t'/\varepsilon_0$ is a small expansion parameter, 
we can regard Eqs.~(\ref{eqn:(9)}) and (\ref{eqn:(10)}) as a 
perturbation to Eqs.~(\ref{eqn:(7)}) and (\ref{eqn:(8)}); 
the latter parts are essentially a spinless fermion Hamiltonian  
with nearest-neighbor hopping. The degeneracy with respect to spin 
can be lifted only after Eqs.~(\ref{eqn:(9)}) and (\ref{eqn:(10)}) are 
introduced.

\begin{figure*}
  \epsfxsize=8.5 truecm
  \centerline{\epsffile{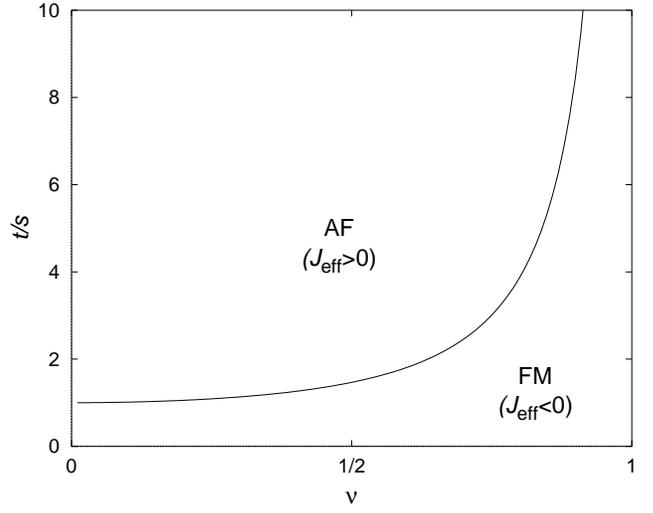}}
  \caption{ 
  For $U=+\infty$ and $t'/\varepsilon_0\ll 1$ the sign of 
  $J_{\rm eff}$ and the nature of the ground
  state is determined by the ratio 
  $t/s$. For  negative $t$ it will always give a ferromagnetic 
  effective coupling.
  }
  \label{fig:tvss}
\end{figure*}

As far as $t'/\varepsilon_0$ is small, we can apply the perturbation 
treatment which was used for the large--$U$ limit of the 
Hubbard model.\cite{ogata} Just like in the latter case the 
ground state wave function 
$\vert \Psi_g\rangle$ is given as a direct product of spinless 
fermion wave function $\vert\Phi_{\rm SF}\rangle$ 
and spin wave function $\vert\chi\rangle$: 
\begin{equation}
 \vert \Psi_g\rangle
 =\vert\Phi_{\rm SF}\rangle \otimes \vert\chi\rangle. 
 \label{eqn:(13)}
\end{equation}
To discuss the ground state of the effective Hamiltonian 
we take for $\vert\Phi_{\rm SF}\rangle$ 
the spinless fermion ground state of 
\begin{equation}
{\cal H}_{\rm SF}=s\sum_{j}(a_{j}^{\dagger}a_{j+1}+{\rm H.c.}), 
\label{eqn:(14)}
\end{equation}
where $a_j$ and $a_j^{\dagger}$ are 
spinless fermion annihilation and 
creation operators respectively. 
$\vert\chi\rangle$ has to be determined to minimize 
the spin-dependent part of the effective Hamiltonian. 
Taking the average of ${\cal H}_{\rm eff}$ 
over $\vert\Phi_{\rm SF}\rangle$ and picking up 
the spin-dependent part, we obtain
\begin{eqnarray}
\langle{\cal H}_{\rm eff}\rangle_{\rm SF}&=&J_{\rm eff}
\sum_{i}{\bf S}_i\cdot{\bf S}_{i+1}, 
\label{eqn:(14a)} \\
J_{\rm eff}&=&-4\Big({t' \over \varepsilon_0}\Big)^2
\Big[s g(\nu)+t f(\nu)\Big]
+4{s^2 \over U}h(\nu) 
\label{eqn:(15)}
\end{eqnarray}
where 
\begin{eqnarray}
g(\nu)&=&\nu^2\Big[1-\Big({\sin(\pi\nu) \over \pi\nu}\Big)^2\Big], 
\label{eqn:(16)}
\\
f(\nu)&=&{1 \over \pi^2}\sin(\pi\nu)\Big[\pi\nu\cos(\pi\nu)
-\sin(\pi\nu)\Big], \label{eqn:(17)} \\
h(\nu)&=&\nu^2\Big(1-{\sin(2\pi\nu) \over 2\pi\nu}\Big).
\label{eqn:(17a)}
\end{eqnarray}
The summation in Eq.~(\ref{eqn:(14a)}) is taken over the 
{\it squeezed} spins.\cite{ogata} 
$\nu$ is twice the density $n$ of electrons. 
For $0\le \nu\le 1$, $g(\nu)$ and $h(\nu)$ are positive, 
while $f(\nu)$ is negative. Therefore, for positive $s$ 
and $t$, a competition occurs among three terms: the 
$s(t'/\varepsilon_0)^2$ term favors ferromagnetism, 
whereas the $t(t'/\varepsilon_0)^2$ and $s^2/U$ terms tend to destabilize it. 
Fig. \ref{fig:tvss} shows the sign of $J_{\rm eff}$. 
For $0\le \nu \le 1$, the ground state is 
ferromagnetic as long as $J_{\rm eff}$ is negative.

  Now let us turn to the excitation spectrum of the one spin flipped state 
in the same parameter region of $J_{\rm eff}\ll t$. The low energy excitation 
of the spin part are the spin waves with momentum $q=2 \pi J/N_e$ and
energy $\varepsilon(q) = J_{\rm eff} (1-\cos q)$. The momentum of
the spin part changes the boundary conditions of the spinless fermions, 
as noted by Woynarovich,\cite{bogyo} 
so that the momenta of spinless fermions are $(L/2) k_j = 2 \pi I_j + q$ with 
$I_j$ integers, and this phase shift appears
in the energy and momentum of the spinless fermion part:
\begin{eqnarray}
  E &=& J_{\rm eff} (1-\cos q) -2 s\sum_{j} \cos \frac{2 \pi I_j + q}{(L/2)},
  \nonumber\\ 
  P &=& \sum_{j} k_j = 2\pi \frac{2}{L} \sum_j I_j + \frac{2 N_e}{L} q.
\end{eqnarray}
In the thermodynamic limit,
if the ground state of $q=0$ (i.e. of the highest spin state) is 
nondegenerate, then the energy $E(P)-E_{\rm FM}$ of the lowest 
lying excitations of the 
system is given by 
\begin{equation}
  \omega(k) = 
 J_{\rm eff} (1-\cos q) + \frac{2}{L}s q^2 \frac{\sin \pi \nu}{\pi}
\end{equation}
where $k=\nu q + P_{\rm FM}$. 
Unlike the insulating case at $n=\case{1}{2}$, 
this spin wave 
excitation is not separated any more by a finite gap from the Stoner 
continuum. We do not know the relevance of this fact to the stability 
of itinerant ferromagnetism in this case.

 Depending on the number of particles, see Table.~\ref{tab:deg}, 
it can easily happen that the 
fully polarized state is degenerate. In this case of
 orbital degeneracy  the energy of the spin wave with momentum 
$k=\nu q \pm |P_{\rm FM}|$ 
becomes
\begin{equation}
  \omega(k) = 
 J_{\rm eff} (1-\cos q) + \frac{2}{L}s (q\mp\pi)^2 \frac{\sin \pi \nu}{\pi}
\end{equation}
where we can clearly identify the competition between the kinetic energy gained
by twisting the boundary conditions and the energy loss due to the spin 
wave. Actually, in the latter case, unless 
$J_{\rm eff}>J_c \propto s \nu \sin \pi\nu$ , 
the minimum energy will occur for some finite $q$, and it prevents the 
highest spin state from being the ground state.

\begin{figure*}
  \epsfxsize=8.5 truecm
  \centerline{\epsffile{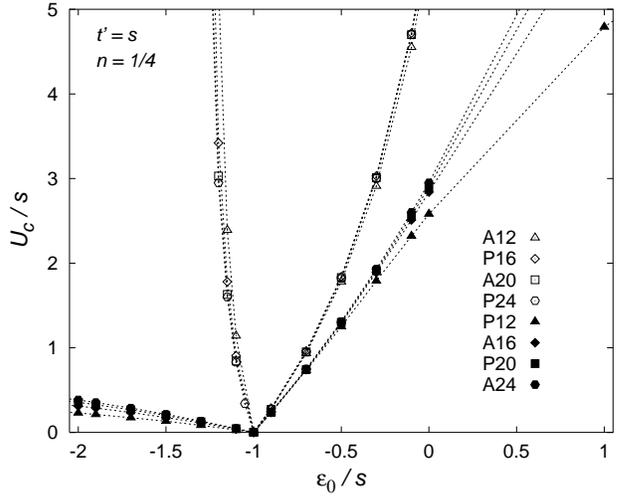}}
  \caption{
  Local stability for $n=1/4$ from exact diagonalization of 12, 16, 20 and 
24 site clusters with periodic (P) or antiperiodic (A) boundary conditions
($t'=s$, $t=0$).
The fully polarized state is degenerate for open symbols. Ferromagnetism is
stable against single spin flip for $U> U_c$.
  }
  \label{fig:locsta2}
\end{figure*}

  We have performed an exact diagonalization study of the model at $n = 1/4$
(i.e. $\nu = 1/2$) to determine the region where the fully 
polarized is stable against spin flip. The results are presented on 
Fig.~\ref{fig:locsta2}. The behavior described above can be clearly seen for
larger values of $U$ and $|\varepsilon_0|$, whereas near the flat band 
for $\varepsilon_0>\varepsilon_{\rm FB}$
the effect of orbital degeneracy is not so important .

\acknowledgements

    We are especially thankful to H. Tasaki for stimulating discussions.
The authors enjoyed interesting discussions with  P. Fazekas,
K. Hallberg, J. S\'olyom, V. Subrahmanyam and X. Zotos.
We also thank IDRIS (Orsay) for allocation of CPU time on 
the C94 and C98 CRAY supercomputers.

\appendix
\section*{Higher order terms in the series expansion}

Expansion of $K_1$ continues as:
\end{multicols}
\widetext
\begin{eqnarray}
K_1^{(4)}&=&
 -{16 s^4 \over U^3} 
 +{4 t'^4 \over \varepsilon_0^2 U} 
 +{8 t'^4 \over \varepsilon_0^2 \left( 2 \varepsilon_0 + U \right) },
\nonumber\\
K_1^{(5)}&=&
 -{12 s^3 t'^2 \over \varepsilon_0^4} 
 +{40 s t'^4 \over \varepsilon_0^4} 
 +{64 s^3 t'^2 \over \varepsilon_0 U^3} 
 +{32 s^3 t'^2 \over \varepsilon_0^2 U^2} 
 +{16 s t'^4 \over \varepsilon_0^3 U}, 
\nonumber\\
K_1^{(6)}&=&
  {36 s^2 t'^4 \over \varepsilon_0^5} 
 +{120 s^6 \over U^5}  
 -{96 s^2 t'^4 \over \varepsilon_0^2 U^3}  
 -{64 s^4 t'^2 \over \varepsilon_0^2 U^3}  
 -{64 s^4 t'^2 \over \varepsilon_0^3 U^2}  
 -{96 s^2 t'^4 \over \varepsilon_0^3 U^2}  
 -{56 s^4 t'^2 \over \varepsilon_0^4 U}  
 -{16 s^2 t'^4 \over \varepsilon_0^4 U}  
 -{16 t'^6 \over \varepsilon_0^4 U} 
 +{8 s^4 t'^2\over \varepsilon_0^4 \left( \varepsilon_0 + U \right) } 
\nonumber\\&& 
 +{16 s^2 t'^4\over \varepsilon_0^2 \left( 2 \varepsilon_0 + U \right)^3 } 
 +{32 s^2 t'^4\over \varepsilon_0^3 \left( 2 \varepsilon_0 + U \right)^2} 
 +{48 s^2 t'^4\over \varepsilon_0^4 \left( 2 \varepsilon_0 + U \right) }  
 -{96 t'^6\over \varepsilon_0^4 \left( 2 \varepsilon_0 + U \right) },
\nonumber\\
K_1^{(7)}&=&
 -{84 \tilde t^7 \over \varepsilon_0^6}  
 -{720\tilde t^7 \over \varepsilon_0 U^5}  
 -{360\tilde t^7 \over \varepsilon_0^2 U^4} 
 +{136\tilde t^7 \over \varepsilon_0^3 U^3} 
 +{236\tilde t^7 \over \varepsilon_0^4 U^2} 
 +{224\tilde t^7 \over \varepsilon_0^5 U} \nonumber\\&& 
 -{28\tilde t^7 \over \varepsilon_0^5 \left( \varepsilon_0 + U \right) }  
 -{32\tilde t^7 \over \varepsilon_0^3 \left( 2\varepsilon_0 + U \right)^3}  
 -{80\tilde t^7 \over \varepsilon_0^4 \left( 2\varepsilon_0 + U \right)^2}  
 -{32\tilde t^7 \over \varepsilon_0^5  \left( 2\varepsilon_0 + U \right)}
\nonumber
\end{eqnarray}
where $\tilde t = t'=s$.
Contribution to $K_2$ starts in 4th order:
\begin{eqnarray}
K_2^{(4)}&=&
  {4 s^4 \over U^3}, 
\nonumber\\
K_2^{(5)}&=&
 -{16 s^3 t'^2 \over \varepsilon_0 U^3}  
 -{8 s^3 t'^2\over \varepsilon_0^2 U^2}, 
\nonumber\\
K_2^{(6)}&=&
  {8 s^2 t'^4 \over \varepsilon_0^5  }  
 -{48 s^6 \over  U^5} 
 +{16 s^4 t'^2 \over \varepsilon_0^2 U^3} 
 +{24 s^2 t'^4 \over \varepsilon_0^2 U^3} 
 +{16 s^4 t'^2 \over \varepsilon_0^3 U^2} 
 +{24 s^2 t'^4 \over \varepsilon_0^3 U^2} 
 +{20 s^4 t'^2 \over \varepsilon_0^4 U} 
 +{4 s^2 t'^4 \over \varepsilon_0^4 U} 
 +{4 s^4 t'^2 \over \varepsilon_0^4 \left( \varepsilon_0 + U \right)} 
 +{4 s^2 t'^4 \over \varepsilon_0^4 \left( 2 \varepsilon_0 + U \right)},
\nonumber\\
K_2^{(7)}&=&
  {288 \tilde t^7\over \varepsilon_0 U^5} 
 +{144 \tilde t^7\over \varepsilon_0 U^4 }  
 -{16 \tilde t^7\over \varepsilon_0^3U^3}  
 -{32 \tilde t^7\over \varepsilon_0^4U^2}  
 -{96 \tilde t^7\over \varepsilon_0^5  U}  
 -{32 \tilde t^7\over \varepsilon_0^5  \left( \varepsilon_0 + U \right)} 
 +{16 \tilde t^7\over \varepsilon_0^5  \left( 2\varepsilon_0 + U \right)},
\nonumber
\end{eqnarray}
\begin{multicols}{2}
\narrowtext
and to $K_3$ in 6th order:
\begin{eqnarray} 
K_3^{(6)} &=& 
  {8 s^6 \over  U^5}, 
\nonumber\\
K_3^{(7)} &=& 
 -{48 \tilde t^7\over \varepsilon_0 U^5}  
 -{24 \tilde t^7\over \varepsilon_0^2 U^4 }  
 -{8 \tilde t^7\over \varepsilon_0^3 U^3}  
 -{12 \tilde t^7\over \varepsilon_0^4U^2}.
  \nonumber
\end{eqnarray}

The coefficients $\bar K_j$ 
for $U,-\varepsilon_0 \gg \tilde t$, where $\tilde t=s=t'$ and we introduced 
the notation $\bar \varepsilon_0 = -\varepsilon_0$, are: 
\begin{eqnarray}
 \bar K_1^{(6)}&=&
  {20\tilde t^6 \over \bar \varepsilon_0^5} 
 +{24\tilde t^6 \over \bar \varepsilon_0^4 U} 
 +{16\tilde t^6 \over \bar \varepsilon_0^3 
      \left( 2\bar \varepsilon_0 + U \right)^2} 
 +{16\tilde t^6 \over \bar \varepsilon_0^4 
      \left( 2\bar \varepsilon_0 + U \right) },
\nonumber\\
\bar K_1^{(7)}&=&
  {72\tilde t^7 \over \bar \varepsilon_0^6} 
 +{32\tilde t^7 \over \bar \varepsilon_0^5 U}  
 -{160\tilde t^7 \over \bar \varepsilon_0^4 
      \left( 2\bar \varepsilon_0 + U \right)^2} 
 +{352\tilde t^7 \over \bar \varepsilon_0^5 
      \left( 2\bar \varepsilon_0 + U \right) },
\nonumber
\end{eqnarray}
and the first contributions to $K_6$ come in 6th order:
\begin{eqnarray}
 \bar K_2^{(6)}&=&
  {4\tilde t^6 \over \bar \varepsilon_0^4 U} 
 +{16\tilde t^6 \over \bar \varepsilon_0^4 
     \left( 2\bar \varepsilon_0 + U \right) },
\nonumber\\
 \bar K_2^{(7)}&=&
 -{8\tilde t^7 \over \bar \varepsilon_0^6}  
 -{8\tilde t^7 \over \bar \varepsilon_0^5 U} 
 +{48\tilde t^7 \over \bar \varepsilon_0^4 
      \left( 2\bar \varepsilon_0 + U \right)^2} 
 -{48\tilde t^7 \over \bar \varepsilon_0^5  
      \left( 2\bar \varepsilon_0 + U \right) }.
\nonumber
\end{eqnarray}

\end{multicols}
\widetext

 \mediumtext
\begin{table}
\caption{ 
   $\varepsilon_{\rm FB}$, constant $C$ and 
critical values of Coulomb repulsion $U$ for some selected values of $t'/s$.
 \label{tab:1}}
\begin{tabular}{cdddddddd}
 &&& \multicolumn{3}{c}{$\rho>0$} &\multicolumn{3}{c}{$\rho<0$}\\
    $t'/s$ & $\varepsilon_{\rm FB}/s$ & $C$
    & $U_D/\rho$ & $U_c/\rho$ & $U_c^{\rm app}/\rho$ 
    & $U_D/|\rho|$ & $U_c/|\rho|$ & $U_c^{\rm app}/|\rho|$\\
\tableline
 4        & 14.   & 0.00154 & 0.830 & 0.831 & 0.223 & 2.402 & 2.414 & 0.271 \\
 2        &  2.   & 0.0136  & 1.047 & 1.077 & 0.707 & 1.851 & 1.945 & 1.\\
$\sqrt{2}$&  0.   & 0.0302  & 1.235 & 1.355 & 1.155 & 1.527 & 1.715 & 1.406\\
1         & -1.   & 0.0557  & 1.476 & 1.860 & 1.789 & 1.277 & 1.555 & 1.505\\
1/2       & -1.75 & 0.149   & 2.169 & 3.88  & 3.88  & 1.025 & 1.294 & 1.294\\
\end{tabular}
\end{table}

\widetext
\begin{multicols}{2}
\narrowtext

\begin{table*}
\caption{ Momentum $P_{\rm FM}$ (with degeneracy) of the fully polarized
state for periodic (PBC) and antiperiodic (APBC) boundary conditions,
 $t=0$.
 \label{tab:deg}}
\begin{tabular}{cccccc}
 && \multicolumn{2}{c}{$\varepsilon_0>\varepsilon_{\rm FB}$} 
    &\multicolumn{2}{c}{$\varepsilon_0<\varepsilon_{\rm FB}$}\\
 $L/2$ & $N_e$ & PBC & APBC & PBC & APBC  \\
\tableline
even & even & $\pm \nu \pi$        & 0                    & $\pm\nu\pi$ & 0 \\
even & odd  & $\pi$                & $\pm (1\!-\!\nu)\pi$ & 0 & $\pm\nu\pi$ \\
odd  & even & $0$                  & $\pm \nu \pi$        & $\pm\nu\pi$ & 0 \\
odd  & odd  & $\pm (1\!-\!\nu)\pi$ & $\pi$                & 0 & $\pm\nu\pi$ \\
\end{tabular}
\end{table*}

\end{multicols}
\end{document}